\documentclass{iaus}
\usepackage{graphicx}

\title[No. 292.~~Molecular Gas, Dust and Star Formation in Galaxies] %% give here short title %%
{Complete Ionisation of the Neutral Gas in High Redshift Radio Galaxies and Quasars}

\author[S. J. Curran \& M. T. Whiting]   %% give here short author list %%
{S. J. Curran$^{1,2}$  \and M. T. Whiting$^3$}

\affiliation{$^1$Sydney Institute for Astronomy, 
The University of Sydney, NSW 2006, Australia \\[\affilskip]
$^2$ARC Centre of Excellence for All-sky Astrophysics (CAASTRO) \\ email: {\tt sjc@physics.usyd.edu.au}\\[\affilskip]
$^3$CSIRO Astronomy and Space Science, PO Box 76, Epping NSW 1710, Australia\\ email: {\tt Matthew.Whiting@csiro.au}
}

\pubyear{2012}
\volume{292}  %% insert here IAU Symposium No.
\pagerange{1--1}
\jname{Molecular Gas, Dust and Star Formation in Galaxies}
\editors{T. Wong \& J. Ott, eds.}

\def \HI {H{\sc \,i}}

\def\lapp{\ifmmode\stackrel{<}{_{\sim}}\else$\stackrel{<}{_{\sim}}$\fi}
\def\gapp{\ifmmode\stackrel{>}{_{\sim}}\else$\stackrel{>}{_{\sim}}$\fi}
\begin{document}

\maketitle

\begin{abstract}
 Cool neutral gas provides the raw material for all star formation in the Universe, and yet, from a survey of the hosts
  of high redshift radio galaxies and quasars, we find a complete dearth of atomic (\HI\ 21-cm) and molecular (OH, CO,
  HCO$^+$ \& HCN) absorption at redshifts $z \gapp3$ (\cite[Curran et al. 2008]{cww+08}). Upon a thorough analysis of the optical photometry, we
  find that all of our targets have ionising ($\lambda\leq 912$ \AA) ultra-violet continuum luminosities of $L_{\rm
    UV}\gtrsim10^{23}$ W Hz$^{-1}$. We therefore attribute this deficit to the traditional optical selection of targets
  biasing surveys towards the most ultra-violet luminous objects, where the intense radiation excites the neutral gas to
  the point where it cannot engage in star formation (\cite[Curran \& Whiting 2010]{cw10}). However, this hypothesis does not explain why there
  is a critical luminosity, rather than a continuum where the detections gradually become fewer and fewer as the
  harshness of the radiation increases. We show that by placing a quasar within a galaxy of gas there is {\em always} a finite
  ultra-violet luminosity above which {\em all} of the gas is ionised.
This demonstrates that
  these galaxies are probably devoid of star-forming material rather than this being at abundances below the sensitivity
  limits of current radio telescopes.

\keywords{galaxies: active --- galaxies: ISM --- radio lines: galaxies --- ultra violet: galaxies} % --- galaxies: high redshift --- cosmology: early universe}
\end{abstract}

For a cloud of hydrogen containing an ionising source, the equilibrium between photoionsation and recombination of
protons and electrons in a nebula is given by
\begin{eqnarray*}
  \int^{\infty}_{\nu_{_{\rm ion}}}\frac{L_{\nu}}{h\nu}\,d{\nu}= 4\pi\int^{r_{\rm ion}}_{0}\,n_{\rm p}\,n_{\rm e}\,\alpha_{A}\,r^2\, dr = 4\pi\,\alpha_{\rm A}\,n_0^2\int^{r_{\rm ion}}_{0}\,e^{-2r/R}\,r^2\, dr,
\label{eq1}
\end{eqnarray*}
for a  gas density $n_{\rm p} = n_{\rm e} = n = n_0\, e^{-r/R}$, where $n_0 \,(= 10$~cm$^{-3}$) is the value at $r = 0$ and $R$ is a scale-length describing the rate of decay of this with
radius. For the observed critical value of $\int^{\infty}_{\nu_{\rm ion}}\ (L_{\nu}/h\nu)\,d\nu = 3\times10^{56}$ ionising
photons sec$^{-1}$ ($L_{\rm  UV}\sim10^{23}$ W Hz$^{-1}$), this gives a scale-length of $R=3$ kpc, which is the value 
found for the \HI\ in the Milky Way, which has an exponential profile with a similar % $n_0 = 0.9\,e^{R_\odot/R} = 13.4$~cm$^{-3}$.
  $n_0 = 13.4$~cm$^{-3}$
(\cite[Kalberla \& Kerp 2009]{kk09}). That is, the observed critical
luminosity is sufficient to ionise all of the gas in a large spiral galaxy,
thus explaining why neutral gas is not detected in high redshift optically selected sources (\cite[Curran \& Whiting 2012]{cw12}).
Therefore, even the SKA is unlikely to detect 21-cm absorption within the host galaxies of these objects.
\vspace*{-3mm} %doesn't work - losing Matt's email address helps


\begin{thebibliography}{}
\bibitem[{Curran \& Whiting(2010)}]{cw10}
{Curran, S.~J. \& Whiting, M.~T.} 2010, \textit{ApJ}, 712, 303

\bibitem[{Curran \& Whiting(2012)}]{cw12}
{Curran, S.~J. \& Whiting, M.~T.} 2012, \textit{ApJ}, in press (arXiv:1204.2881)

\bibitem[Curran \etal\ (2008)]{cww+08}
{Curran, S.~J., Whiting, M.~T., Wiklind, T., Webb, J.~K., Murphy, M.~T., \&
  Purcell, C.~R.} 2008, \textit{MNRAS}, 391, 765

\bibitem[{{Kalberla} \& {Kerp}(2009)}]{kk09}
{Kalberla}, P.~M.~W. \& {Kerp}, J. 2009, \textit{ARAA}, 47, 27

\end{thebibliography}
\end{document}